\documentclass[%
 aip,
 amsmath,amssymb,
 reprint,%
 groupedaddress,
]{revtex4-1}

\usepackage{graphicx}
\usepackage{dcolumn}
\usepackage{bm}

\usepackage[utf8]{inputenc}
\usepackage[T1]{fontenc}
\usepackage{mathptmx}
\usepackage{etoolbox}

\makeatletter
\def\@email#1#2{%
 \endgroup
 \patchcmd{\titleblock@produce}
  {\frontmatter@RRAPformat}
  {\frontmatter@RRAPformat{\produce@RRAP{*#1\href{mailto:#2}{#2}}}\frontmatter@RRAPformat}
  {}{}
}%
\makeatother

\usepackage[colorlinks=true,linkcolor=cyan,urlcolor=cyan,citecolor=cyan]{hyperref}

\begin{document}

\preprint{AIP/123-QED}

\title[Impact-induced viscoelastic bungee-jumper jets with uniform extension and stress]{Impact-induced viscoelastic bungee-jumper jets with uniform extension and stress}
\author{Kyota Kamamoto}
\author{Asuka Hosokawa}
\author{Yoshiyuki Tagawa} 
 \email[Author to whom correspondence should be addressed: ]{tagawayo@cc.tuat.ac.jp}
 \altaffiliation{https://sites.google.com/view/tagawalabjp}
\affiliation{Department of Mechanical Systems Engineering, Tokyo University of Agriculture and Technology, Koganei, Tokyo 184–8588, Japan.}

\date{\today}

\begin{abstract}
We investigate the dynamics of a ``bungee-jumper'' jet induced by an impulsive force, which retracts after reaching its peak extension. 
Despite the strongly extensional and highly nonequilibrium nature of this motion, the jet exhibits simple and uniform rheological responses. 
To elucidate its extensional behavior in a highly extensional regime quantified by large Deborah and Reynolds numbers ($De \approx 2.1\times10^{1}$–$3.3\times10^{3}$, $Re \approx 2.8\times10^{1}$–$4.6\times10^{2}$), we use high-speed velocimetry and polarization-based stress imaging to measure the spatial distribution of velocity and stress throughout jets made of dilute polyethylene oxide (PEO) solutions. 
The bungee-jumper jets are found to exhibit two uniform characteristics despite the extreme $De$ conditions: a consistent spatial distribution of the extensional rate and a nearly uniform stress distribution during the jetting motion. 
These uniformities indicate that the seemingly complex jet dynamics can in fact be effectively represented using a constitutive model with spatially uniform coefficients. 
Comparison of several viscoelastic models shows that the Voigt model provides the best agreement with the measured dynamics, while the single-spring model captures the essential behavior when elasticity dominates.
\end{abstract}

\maketitle

In recent years,  next-generation printing technologies such as three-dimensional printing,\cite{truby2016, crisostomo2021, cheng2022} bio-printing,\cite{duan2017, noor2019, ng2024} and food printing\cite{mantihal2020, pant2021, demei2022} have been extensively investigated. Depending on their specific application, these technologies use a various different types of fluids, among which non-Newtonian fluids such as molten resins and polymer solutions are widely used in the manufacturing industry.\cite{castrejon2013, tamay2019}
For example, in inkjet printing, one method for preventing undesirable satellite droplets is to add polymers to the ink.\cite{christanti2002}
The properties and behavior of such viscoelastic materials have been extensively studied for many years.\cite{colby2010, subbotin2023, sari2024, oratis2025}
Although there have been many studies\cite{mun1998, de2004, xu2019, antonopoulou2020, zhao2021, balasubramanian2024, wang2024, dixit2025, gaillard2025} of liquids whose viscoelastic properties have minor effects, further investigations are needed for cases where the viscoelastic properties are dominant.
In particular, “bungee-jumper jets,” which retract after reaching their maximum extension, have been observed when  viscoelastic effects play a major role.\cite{bazilevskii2005, hoath2009, morrison2010, turkoz2018, franco2021,torres2022}
Such retraction-dominant behavior provides rare access to strongly extensional, highly nonequilibrium rheological responses of complex fluids—conditions that are difficult to realize or measure with conventional rheometric tools.

One method for ejecting liquids at high velocity is the focused liquid jet technique, which uses a laser\cite{tagawa2012, peters2013, delrot2016, delrot2018} or an impact\cite{antkowiak2007, kiyama2016, onuki2018, yukisada2018, kamamoto2021ouzo, kurihara2025} to generate the jet. 
The ejection of polymers by this technique has  potential application in next-generation printing technology, and its study will also provide greater understanding of the extensional rheology of polymers and the effects of their viscoelastic properties.
Impact-induced focused jets are particularly advantageous in this context, because they are associated with extremely high strain rates and high elastic stresses, enabling the study of viscoelastic jet dynamics in regimes that cannot be reached using standard jetting approaches.

The objective of the present study is to elucidate the influence of viscoelastic properties on bungee-jumper jets.
To do this, we vary the controlling parameters in our experiment, namely, the molecular weight of the polymer solution (which is related to the viscoelastic properties of the liquid) and the initial velocity of the fluid.
We measure the spatial distribution of velocity and stress inside the jet.
We then discuss viscoelastic models that can describe  bungee-jumper jet behavior in this extreme regime, showing that the seemingly complex dynamics can in fact be captured by simple constitutive representations with spatially uniform coefficients.

\begin{figure}[!t]
  \includegraphics[trim={0cm 0cm 0cm 0cm}, clip,width=1.0\columnwidth]{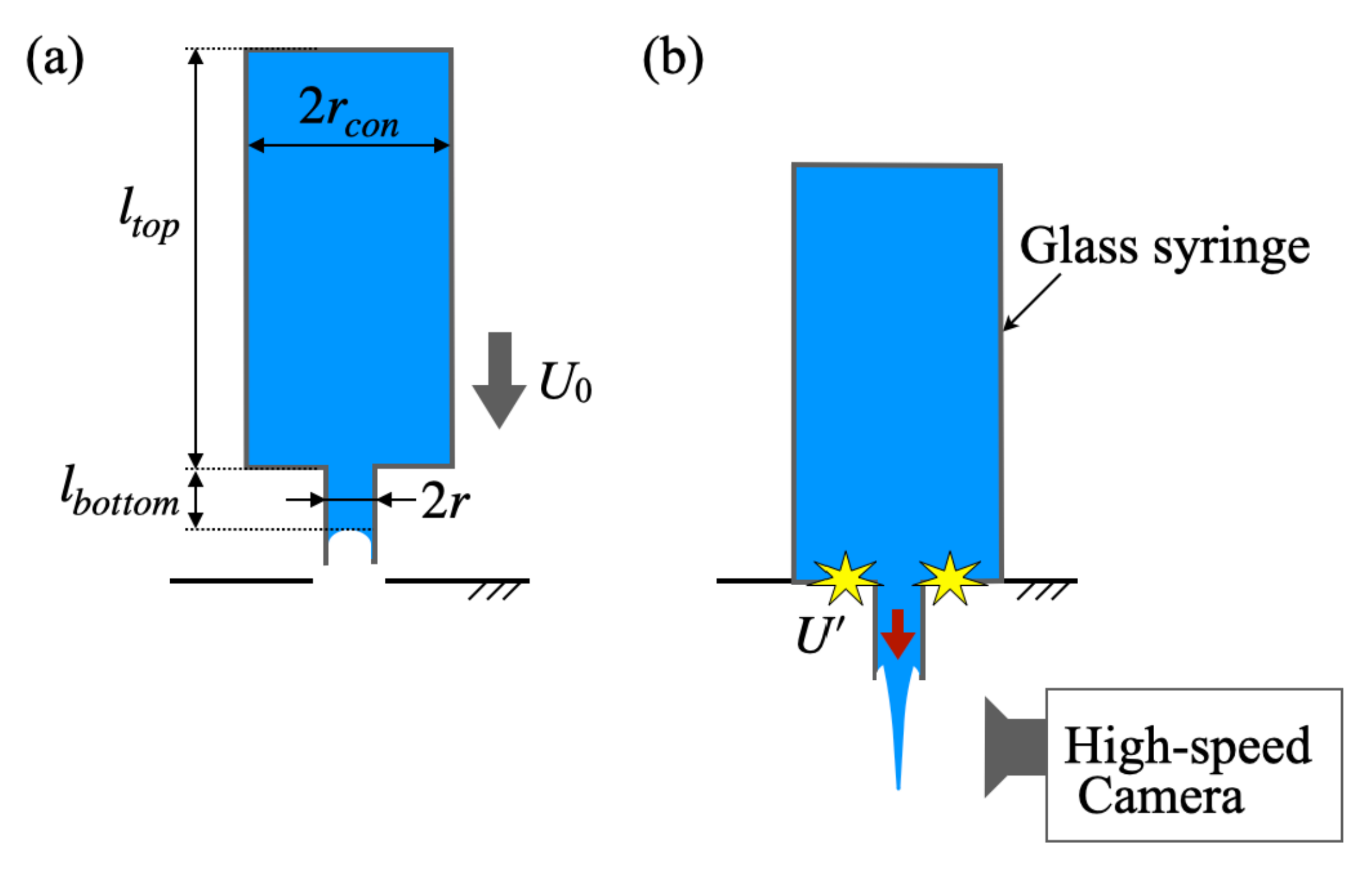}
  \caption{
  Impact-induced liquid-jet generation  using a glass syringe:\cite{kamamoto2021} 
  (a) before impact; (b) just after impact. 
  An impulsive acceleration applied to the container produces a highly focused liquid jet with an extremely large extensional rate, enabling access to flow regimes that are difficult to achieve with conventional jetting or rheometric techniques.
  }
  \label{fig:Setup}
\end{figure}

\begin{figure*}[!t]
  \includegraphics[trim={0cm 0cm 0cm 0cm}, clip,width= 2.0\columnwidth]{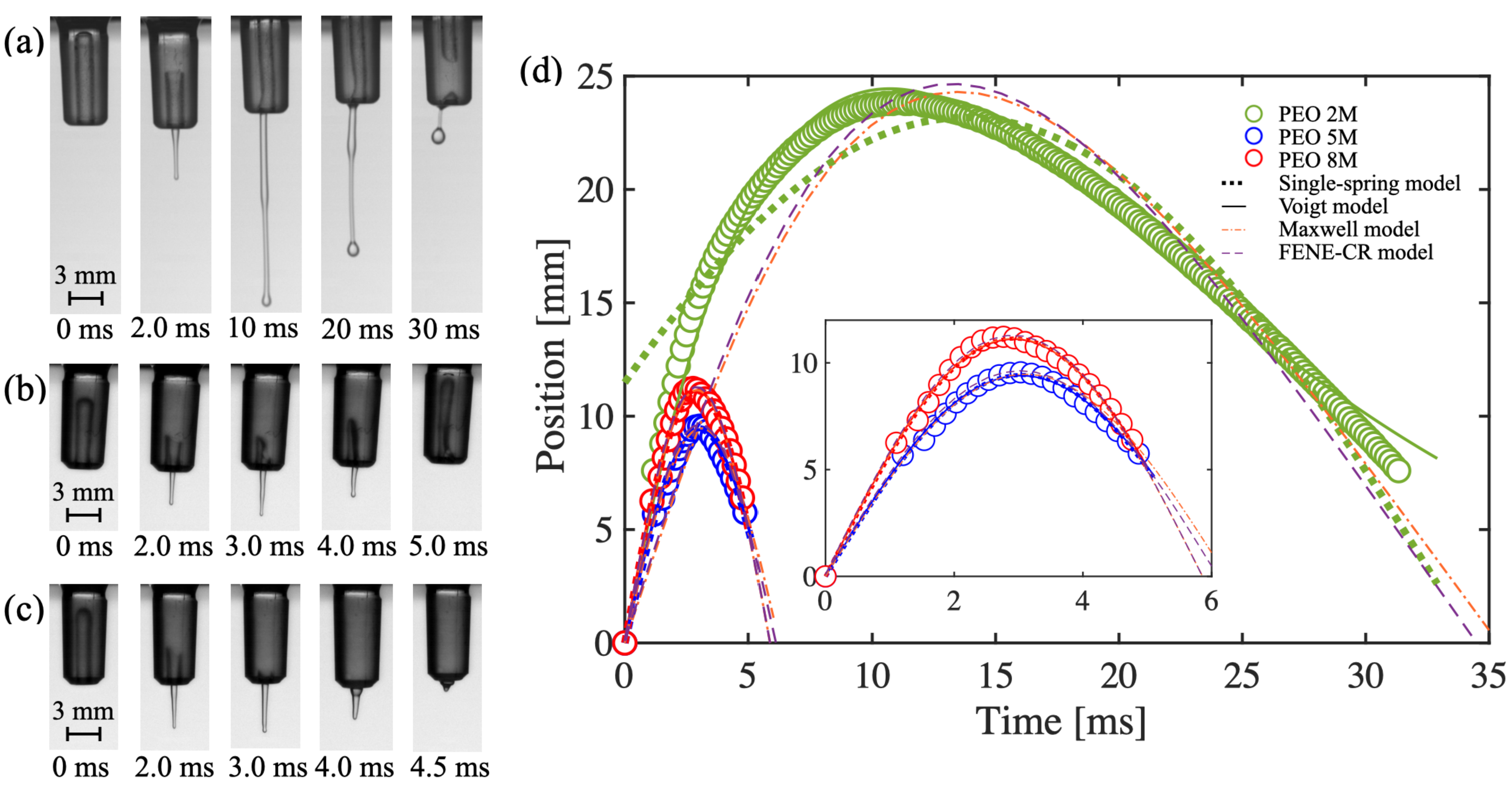}
  \caption{
  Formation and dynamics of bungee-jumper jets for different PEO solutions. 
  (a), (b), and (c) Jet shapes for 2M, 5M, and 8M PEO solutions at similar initial velocities ($U' \approx 5~\mathrm{m/s}$). 
  In all cases, the jet reaches a maximum extension before retracting toward the nozzle, but the extent and timing of retraction differ markedly across molecular weights (see Movie S1 in the supplementary material). 
  (d) Time evolution of jet-tip displacement. 
  Circles denote experimental data; green dotted and green solid curves represent fits by the single-spring and Voigt models, respectively. 
  Orange dot-dashed and purple dashed curves show approximate predictions by the Maxwell and FENE–CR models. 
  Across the range of viscoelasticities examined, the Voigt model provides the most consistent description, capturing both weak- and strong-elasticity cases, whereas the single-spring model is valid only when elasticity dominates (5M and 8M).
  }
  \label{fig:Result}
\end{figure*}

The experimental setup is shown in Fig.~\ref{fig:Setup}.\cite{kamamoto2021}
To eject impact-induced high-velocity liquid jets, we employ a liquid-jet generation technique using a glass syringe.
In this setup, the nozzle radius, container radius, liquid column heights, and impact velocity are set to 
$r = 0.6~\mathrm{mm}$, $r_{\rm con} = 6.2~\mathrm{mm}$, $l_{\rm top} \approx 33\text{–}60~\mathrm{mm}$, 
$l_{\rm bottom} \approx 1.0\text{–}8.2~\mathrm{mm}$, and $U_0 \approx 0.39\text{–}0.81~\mathrm{m/s}$.
A previous study\cite{kamamoto2021} reported that the initial jet velocity at the nozzle, $U'$, can be estimated as
\begin{equation}
U'
= 
\frac{-(2 l_{\rm bottom} + r) + \sqrt{(2 l_{\rm bottom} + r)^2 + 8 r l_{\rm top}}}{2 r}
\, U_0.
\end{equation}

To observe the jet behavior, we use a high-speed camera (FASTCAM SA-X, Photron; frame rate $30\,000$~fps) and a white LED backlight.
We also conduct velocity-distribution\cite{van2014} and polarization\cite{miyazaki2021, yokoyama2023, nakamine2024, worby2024} measurements.
The velocity distribution is obtained by tracking the displacement of segmented volume elements between successive frames.
For polarization measurements, the stress is estimated from the measured phase retardation.\cite{miyazaki2021}
In this experiment, we use a high-speed polarization camera (CRYSTA P1-P, Photron; frame rate $20\,000$~fps) and cellulose nanocrystals (CNC-HS-FD, 0.5~wt.\,\%) dissolved in polyethylene oxide (PEO) 5M solution (1.0~wt.\,\%) as the birefringent probe.

We use PEO solutions of concentration 1.0~wt.\,\%.
The Deborah number spans the range $De \approx 21\text{–}3.3\times10^3$, based on $De = \lambda / t_\mu$ with $t_\mu = \mu_\infty r / \sigma$.\cite{clasen2006}
Similarly, the capillary number spans the range $Ca \approx 1.2\text{–}5.8$, the Reynolds number the range $Re \approx 28\text{–}4.6\times10^2$, 
and the Ohnesorge number the range $Oh \approx 0.079\text{–}0.34$.

Figures~\ref{fig:Result}(a)–\ref{fig:Result}(c) show the formation of bungee-jumper jets for all three PEO solutions at similar initial liquid velocities ($U' \approx 5~\mathrm{m/s}$).
In all three cases, the jet reaches its maximum extension and is then pulled back toward the nozzle.
Despite being driven under similar impact conditions, the three jets exhibit markedly different retraction behaviors, highlighting that viscoelasticity—rather than inertia alone—governs the subsequent dynamics.
To understand the jet behavior in more detail, we investigate the displacement of the jet tip as a function of time [Fig.~\ref{fig:Result}(d)].
The maximum extension lengths are $23.8~\mathrm{mm}$ for the 2M solution, $9.6~\mathrm{mm}$ for 5M, and $11.2~\mathrm{mm}$ for 8M.
In the case of the 2M solution, the jet exhibits more than twice the extension obtained for the 5M and 8M solutions.

\begin{figure}[!t]
  \includegraphics[trim={0cm 0cm 0cm 0cm}, clip,width=1.0\columnwidth]{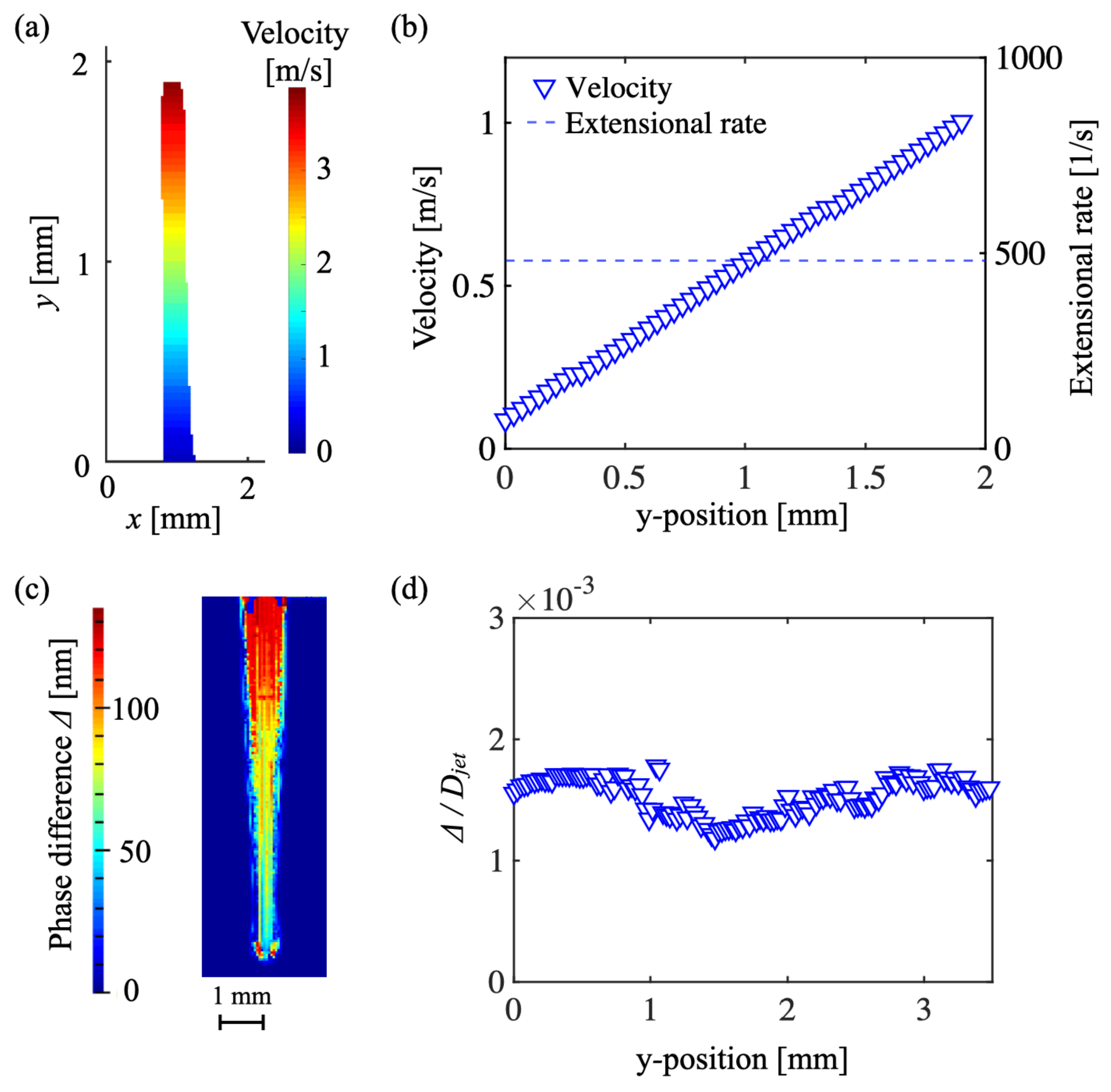}
  \caption{
  Velocity and stress fields of the bungee-jumper jet (PEO 5M) at its maximum extension. 
  (a) Velocity field obtained from high-speed velocimetry and (b) corresponding axial velocity profiles, showing a nearly linear variation with $y$, indicative of a uniform extensional rate.  
  (c) Phase retardation $\Delta$ measured using high-speed polarization imaging and (d) birefringence $\Delta/D_{\mathrm{jet}}$ after correction for the jet diameter. 
  Because phase retardation is related to the path-integrated normal stress through the stress–optic law,\cite{miyazaki2021, yokoyama2023, nakamine2024, worby2024} the nearly uniform birefringence profile indicates that the extensional stress is also nearly uniform. 
  Together, these results demonstrate two key uniformities of bungee-jumper jets: (i) a uniform extensional rate and (ii) a nearly uniform extensional stress along the jet.
  }
  \label{fig:Vel_Stress}
\end{figure}

To clarify why jets with different viscoelasticities retract so differently, we next examine the internal kinematics of the jets through their velocity fields.
The velocity distribution (axially averaged velocity) is shown in Figs.~\ref{fig:Vel_Stress}(a) and~\ref{fig:Vel_Stress}(b).
The velocities are obtained from images taken when the jet is first ejected from the nozzle tip and $0.5~\mathrm{ms}$ later.
These images correspond to the data presented in Fig.~\ref{fig:Result}(d) for the  5M PEO solution.
These data show the distribution of velocity during the extension of the jet, and we have confirmed that similar profiles are observed at other times as well.
The velocity of the bungee-jumper jet is linearly distributed as a function of the $y$ position, indicating that the jet experiences a nearly uniform extensional rate during its flight.
This linear profile is a remarkable result, given the large Deborah numbers and the strongly extensional, highly nonequilibrium nature of the flow.

Complementing the velocity measurements, we also quantify the internal stress distribution to determine whether the uniformity observed in the extensional rate is reflected in the material response.
Because the phase retardation measured in polarization imaging is related to the path-integrated normal stress through the stress–optic law,\cite{miyazaki2021, yokoyama2023, nakamine2024} jets exhibiting identical corrected retardation $\Delta/D_{\rm jet}$ can therefore be regarded as experiencing comparable extensional stress.
Figures~\ref{fig:Vel_Stress}(c) and~\ref{fig:Vel_Stress}(d) show the results of polarization measurement  at the maximum extension.
In this experiment, we observe the jet under the assumption of axisymmetry, and so  the optical depth can be regarded as the jet diameter.
Since the depth values are integrated in the polarization measurement, it is necessary to consider the jet diameter $D_{\rm jet}$ to calculate the birefringence.
Therefore, we plot the birefringence $\Delta / D_{\rm jet}$ in Fig.~\ref{fig:Vel_Stress}(d).
Owing to the limitations of the camera resolution, $\Delta / D_{\rm jet}$ at the very tip cannot be determined accurately, and values are therefore plotted only for $y < 3.5~\mathrm{mm}$.
Remarkably, the resulting birefringence profile shows that the stress is also nearly uniform along the jet, revealing a second unexpected uniformity in this extreme extensional regime.
Hence, the bungee-jumper jet has two distinctive characteristics: (i) the extensional rate is uniform and (ii) the stress is almost constant along the jet during its motion.

The simultaneous uniformity of both extensional rate and stress suggests that the dynamics, although apparently complex, may be effectively described by a simple constitutive model with spatially uniform coefficients.
Our experimental data on velocity and stress distributions therefore motivate a reduced description in terms of a uniform viscoelastic element.
To test this idea, we evaluate and compare several models: the single-spring model, the Voigt model, the Maxwell model, and the FENE--CR model.

Among these candidates, the Voigt model provides the most consistent description of the observed jet dynamics across all PEO solutions.
As  bungee-jumper jets occur under conditions where $De = \lambda / t_{\mu} > 1$, we adopt the Voigt model,\cite{Sarban2011} in which the force is given by $F = -Gx - c\,dx/dt$, where $G$ is the elastic modulus, $c$ is the viscosity coefficient, $x$ is the displacement, and $t$ is time.
The equation of motion of the jet can then be written as
\begin{equation}
\frac{d^{2}x}{dt^{2}} + \frac{c}{m}\frac{dx}{dt} + \frac{G}{m}x = 0,
\label{eq:Voigt_1}
\end{equation}
where $m$ is the mass.
Under the conditions under which bungee-jumper jets occur, elasticity is greater than viscosity, and so the model reduces to
\begin{equation}
x(t) = A e^{-\alpha t}\sin(B t),
\label{eq:Voigt}
\end{equation}
where $A$ is the extension coefficient, $B = \sqrt{G/m}$ corresponds to the angular frequency, and $\alpha$ is the damping rate.
The coefficient $A$ is the amplitude of the bungee-jumper jet behavior, which is related to the maximum extension of the jet.
The fitting results obtained for the Voigt model are shown by the solid line in Fig.~\ref{fig:Result}(d).
The respective coefficients are $A = 147$, $B = 0.0406$~rad/s, and $\alpha = 0.0873$ for the 2M solution. $A = 9.76$, $B = 0.509$~rad/s, and $\alpha = 0.0124$  for the  5M solution, and $A = 11.5$, $B = 0.530$~rad/s, and $\alpha = 0.0108$  for the 8M solution.
The Voigt model is able to approximate the bungee-jumper jet behavior for all three solutions.
In other words, a relatively simple Voigt element with uniform coefficients can capture the essential viscoelastic behavior in this impact-induced jetting problem.

By contrast, the single-spring model captures the jet behavior only in the strongly elastic regime (5M and 8M solutions), reflecting that inertia–viscosity coupling is essential for the 2M case [see the green dotted curve in Fig.~\ref{fig:Result}(d)].
Here, the single-spring model is derived by treating the bungee-jumper jet behavior as a perfectly elastic uniform body (a single spring).
Under the assumption that the force is given by $F = -Gx$, the equation of motion can be written as
\begin{equation}
\frac{d^{2}x}{dt^{2}} + \frac{G}{m}x = 0.
\label{eq:motion}
\end{equation}
The general solution of this ordinary differential equation, obtained under the condition that $G/m > 0$, is
\begin{equation}
x(t) = A \sin(B t + \phi),
\label{eq:x_ss}
\end{equation}
where $A$ is the extension coefficient, $B = \sqrt{G/m}$ corresponds to the angular frequency, and $\phi$ is the initial phase.
The coefficients $A$ and $B$ have the same physical meaning as in the Voigt model.
As mentioned above, the single-spring model cannot describe the 2M solution, which has relatively weak elastic properties, because it does not account for viscous dissipation.

For completeness, we also compare the measured dynamics with the Maxwell and FENE--CR models.\cite{chilcott1988, hoath2012, mcilroy2013}
Since the force in the Maxwell model is given by $F = \lambda\,dF/dt - c\,dx/dt$, the equation of motion can be written as
\begin{equation}
\frac{d^{2}x}{dt^{2}} = \frac{G}{m} x e^{-t/\lambda},
\label{eq:maxwell}
\end{equation}
and the equation of motion in the FENE--CR model is 
\begin{equation}
\frac{d^{2}x}{dt^{2}} = -\frac{K}{\rho}\!
\left(
\frac{3\mu}{x^{2}} \frac{dx}{dt} 
+ \frac{Gx}{r^{2}} e^{-t/\lambda}
\right),
\label{eq:FENE}
\end{equation}
where $K$ is the ratio of the jet-tip radius to the radius of the device.
Using Eqs.~(\ref{eq:maxwell}) and~(\ref{eq:FENE}), approximate solutions are calculated for given initial conditions (position and velocity).
Results for each model are shown as orange dot-dashed and purple dashed curved in Fig.~\ref{fig:Result}(d).
These results indicate that both models can approximately predict the bungee-jumper jet behavior, but not as accurately as the Voigt model. See the supplementary material for more details on the properties of the liquids used in the experiments and the approximation accuracy of
each model.

\begin{figure}[!t]
  \includegraphics[trim={0cm 0cm 0cm 0cm}, clip,width=0.9\columnwidth]{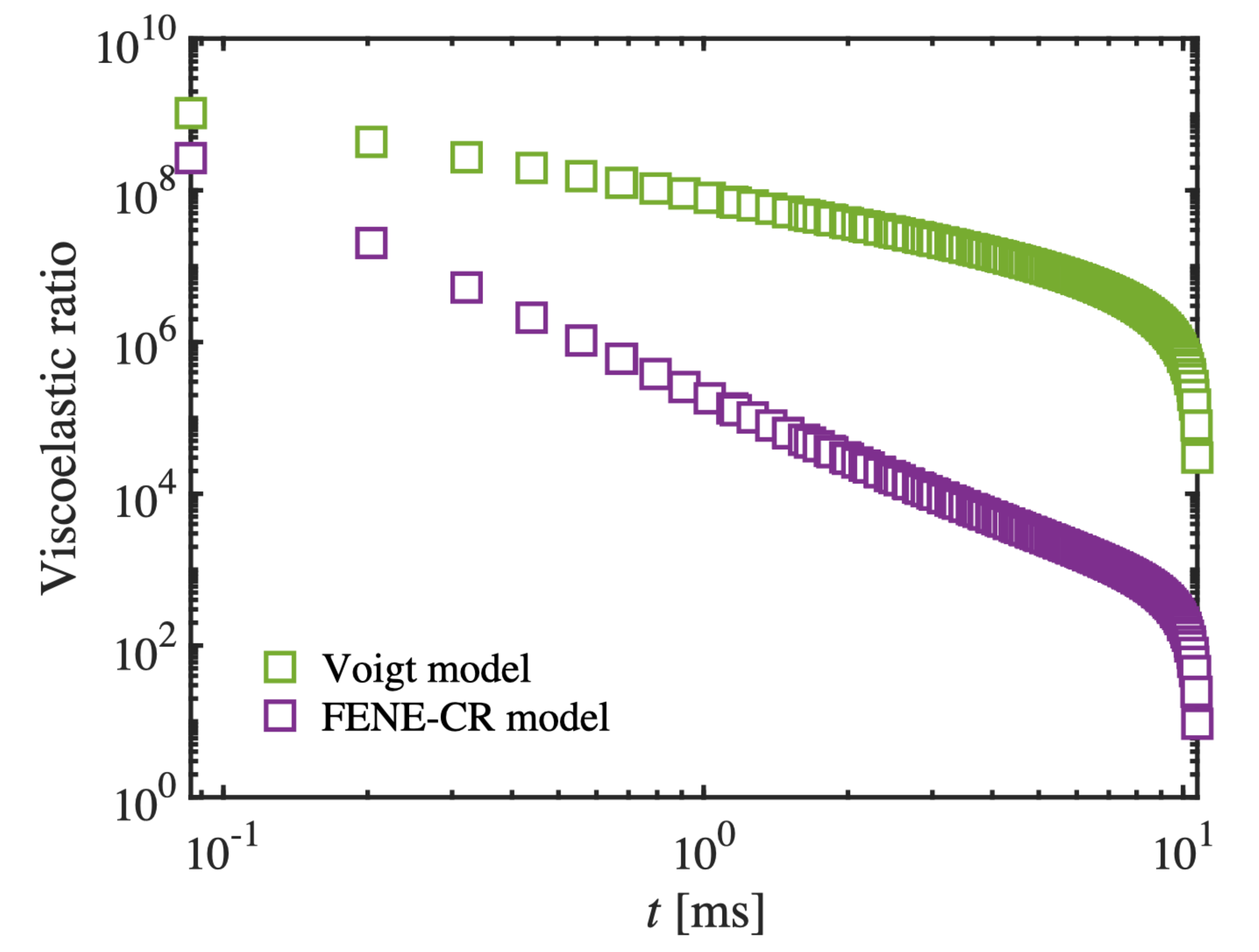}
  \caption{
  Comparison of viscous and elastic force contributions for the 2M PEO solution in the Voigt and FENE–CR models. 
  The viscoelastic ratio quantifies the relative magnitude of viscous to elastic terms in each model. 
  The Voigt model exhibits a substantially larger viscous contribution than the FENE–CR model, explaining its better agreement with experiments, even in regimes where elasticity remains significant. 
  This trend supports the interpretation that viscosity plays an essential role in the extensional dynamics of bungee-jumper jets under the high-$De$, moderate-$Re$ conditions of this study.
  }
  \label{fig:trend}
\end{figure}

To further clarify why the Voigt model performs better than the other models, we calculate the trends of the viscous and elastic terms in each model for the 2M solution [Fig.~\ref{fig:trend}].
The viscous and elastic contributions in the Voigt model are $({c}/{m})\,dx/dt$ ($c/m = 147$) and $(G/m)x$ (with $\sqrt{G/m} = 0.0406$), respectively, while those in the FENE--CR model are $(3K\mu_{0}/\rho x^{2})\,dx/dt$ and $(KG/\rho r^{2})x e^{-t/\lambda}$ (with $KG/\rho r^{2} = 0.033$).
Figure~\ref{fig:trend} shows that the relative magnitude of viscous to elastic terms in each model. The Voigt model is more strongly influenced by the viscous term than the FENE--CR model.
This indicates that the Voigt model can represent the effect of viscosity in the jet dynamics more accurately.
This suggests that the present extensional behavior can be described by neglecting the dashpot on the solvent branch in the Oldroyd-B model, which is commonly used for viscoelastic filaments and liquid jets.\cite{bhat2010, turkoz2018-2, turkoz2021}
In cases where elasticity is dominant, the simpler single-spring model also proves to be a reliable representation.

In conclusion, we have investigated  bungee-jumper jet behavior to elucidate the role of extensional viscoelasticity under strongly extensional, high-$De$ conditions.
In our experiments, impact-induced viscoelastic jets were observed to reach a maximum extension before being pulled back toward the nozzle.
By quantifying the distributions of velocity and stress inside the jets, we found two prominent and unexpected characteristics: a nearly uniform spatial distribution of the extensional rate and an almost uniform stress distribution during the jetting motion.
These uniformities motivated a reduced description in which the jet dynamics are modeled as a uniform viscoelastic element.
Applying several viscoelastic models to the experimental data, we showed that the single-spring model, which considers only elasticity, can describe  bungee-jumper jet behavior for liquids with higher elasticity (5M and 8M solutions), whereas the Voigt model, which incorporates both liquid viscosity and elasticity, can describe the jet behavior even in the weak-elasticity case (2M solution).
Furthermore, under the conditions of this study ($De \approx 21\text{–}3.3\times10^3$, $Re \approx 28\text{–}4.6\times10^2$), the Voigt model consistently provides the best description of impact-induced focused viscoelastic bungee-jumper jets among the models considered, indicating that viscosity remains important even in elasticity-dominant regimes.
These findings demonstrate that the highly nonequilibrium extensional dynamics of viscoelastic jets—despite their apparent complexity—can be reduced to a simple representation with spatially uniform coefficients, providing a concise physical picture of the transition between flight and elastic retraction in complex-fluid jetting.

\bigskip

This work was funded by the Japan Society for the Promotion of Science (Grant Nos.~20H00223, 20H00222, 20K20972, and 24H00289) and the Japan Science and Technology Agency (Grant Nos.~PRESTO JPMJPR21O5 and SBIR JPMJST2355). 
The authors would like to thank Dr. Andres Fernando Franco Gomez and Dr. Prasad Sonar for helpful discussions and comments.

\section*{Author declarations}
\subsection*{Conflict of Interest}
The authors have no conflicts to disclose.

\subsection*{Author Contributions}
All authors approved the final submitted draft.\\

\noindent{\textbf{Kyota Kamamoto}: Conceptualization (equal); Data curation (lead); Formal Analysis (equal); Methodology (lead); Software (equal); Validation (lead); Visualization (lead); Writing/Original Draft Preparation (lead); Writing/Review\&Editing (equal).
\textbf{Asuka Hosokawa}: Data curation (supporting); Formal Analysis (supporting); Software (equal); Visualization (supporting); Writing/Review\&Editing (supporting). 
\textbf{Yoshiyuki Tagawa}: Conceptualization (equal); Formal Analysis (equal); Funding Acquisition (lead); Project Administration(lead); Resources (lead); Supervision (lead); Writing/Review\&Editing (equal).}  

\section*{Data Availability}
The data that support the findings of this study are available from the corresponding author upon reasonable request.

\bibliography{Bungeejet_2}

@PREAMBLE{
"\providecommand{\noopsort}[1]{}" 
# "\providecommand{\singleletter}[1]{#1}%" 
}

@article{antkowiak2007,
title={Short-term dynamics of a density interface following an impact},
author={Antkowiak, Arnaud and Bremond, Nicolas and Le Diz{\`e}s, St{\'e}phane and Villermaux, Emmanuel},
journal={J. Fluid Mech.},
volume={577},
pages={241--250},
year={2007},
publisher={Cambridge University Press}
}

@article{peters2013,
title={Highly focused supersonic microjets: Numerical simulations},
author={Peters, Ivo R and Tagawa, Yoshiyuki and Oudalov, Nikolai and Sun, Chao and Prosperetti, Andrea and Lohse, Detlef and van der Meer, Devaraj},
journal={J. Fluid Mech.},
volume={719},
pages={587--605},
year={2013},
publisher={Cambridge University Press}
}

@article{onuki2018,
title={Microjet generator for highly viscous fluids},
author={Onuki, Hajime and Oi, Yuto and Tagawa, Yoshiyuki},
journal={Phys. Rev. Appl.},
volume={9},
number={1},
pages={014035},
year={2018},
publisher={APS}
}

@article{delrot2016,
title={Inkjet printing of viscous monodisperse microdroplets by laser-induced flow focusing},
author={Delrot, Paul and Modestino, Miguel A and Gallaire, Fran{\c{c}}ois and Psaltis, Demetri and Moser, Christophe},
journal={Phys. Rev. Appl.},
volume={6},
number={2},
pages={024003},
year={2016},
publisher={APS}
}

@article{tagawa2012,
title={Highly focused supersonic microjets},
author={Tagawa, Yoshiyuki and Oudalov, Nikolai and Visser, Claas Willem and Peters, Ivo R and van der Meer, Devaraj and Sun, Chao and Prosperetti, Andrea and Lohse, Detlef},
journal={Phys. Rev. X},
volume={2},
number={3},
pages={031002},
year={2012},
publisher={APS}
}

@article{kiyama2016,
title={Effects of a water hammer and cavitation on jet formation in a test tube},
author={Kiyama, Akihito and Tagawa, Yoshiyuki and Ando, Keita and Kameda, Masaharu},
journal={J. Fluid Mech.},
volume={787},
pages={224--236},
year={2016},
publisher={Cambridge University Press}
}

@article{colby2010,
title={Structure and linear viscoelasticity of flexible polymer solutions: Comparison of polyelectrolyte and neutral polymer solutions},
author={Colby, Ralph H},
journal={Rheol. Acta},
volume={49},
pages={425--442},
year={2010},
publisher={Springer}
}

@article{mun1998,
title={The effects of polymer concentration and molecular weight on the breakup of laminar capillary jets},
author={Mun, Robert P and Byars, Jeffrey A and Boger, David V},
journal={J. Non-Newtonian Fluid Mech.},
volume={74},
number={1-3},
pages={285--297},
year={1998},
publisher={Elsevier}
}

@article{de2004,
title={Inkjet printing of polymers: state of the art and future developments},
author={De Gans, B-J and Duineveld, Paul C and Schubert, Ulrich S},
journal={Adv. Mater.},
volume={16},
number={3},
pages={203--213},
year={2004},
publisher={Wiley Online Library}
}

@article{xu2019,
title={Phase diagram of pinch-off behaviors during drop-on-demand inkjetting of alginate solutions},
author={Xu, Changxue and Zhang, Zhengyi and Huang, Yong and Xu, Heqi},
journal={J. Manuf. Sci. Eng.},
volume={141},
number={9},
pages={091013},
year={2019},
publisher={American Society of Mechanical Engineers}
}

@article{kamamoto2021,
title={Drop-on-demand painting of highly viscous liquids},
author={Kamamoto, Kyota and Onuki, Hajime and Tagawa, Yoshiyuki},
journal={Flow},
volume={1},
pages={E6},
year={2021},
publisher={Cambridge University Press}
}

@article{castrejon2013,
title={Future, opportunities and challenges of inkjet technologies},
author={Castrejon-Pita, Jose Rafael and Baxter, WRS and Morgan, J and Temple, S and Martin, GD and Hutchings, Ian M},
journal={At. Sprays},
volume={23},
number={6},
year={2013},
publisher={Begel House Inc.}
}

@article{tamay2019,
title={{3D} and {4D} printing of polymers for tissue engineering applications},
author={Tamay, Dilara Goksu and Dursun Usal, Tugba and Alagoz, Ayse Selcen and Yucel, Deniz and Hasirci, Nesrin and Hasirci, Vasif},
journal={Front. Bioeng. Biotechnol.},
volume={7},
pages={164},
year={2019},
publisher={Frontiers Media SA}
}

@article{truby2016,
title={Printing soft matter in three dimensions},
author={Truby, Ryan L and Lewis, Jennifer A},
journal={Nature},
volume={540},
number={7633},
pages={371--378},
year={2016},
publisher={Nature Publishing Group UK London}
}

@article{duan2017,
title={State-of-the-art review of {{3D}} bioprinting for cardiovascular tissue engineering},
author={Duan, Bin},
journal={Ann. Biomed. Eng.},
volume={45},
pages={195--209},
year={2017},
publisher={Springer}
}

@article{crisostomo2021,
title={{3D} printing applications in agriculture, food processing, and environmental protection and monitoring},
author={Crisostomo, Jan Lloyd Buenaventura and Dizon, John Ryan Cortez},
journal={Adv. Sustain. Sci. Eng. Technol.},
volume={3},
number={2},
pages={372312},
year={2021},
publisher={Universitas PGRI Semarang}
}

@article{pant2021,
title={{3D} food printing of fresh vegetables using food hydrocolloids for dysphagic patients},
author={Pant, Aakanksha and Lee, Amelia Yilin and Karyappa, Rahul and Lee, Cheng Pau and An, Jia and Hashimoto, Michinao and Tan, U-Xuan and Wong, Gladys and Chua, Chee Kai and Zhang, Yi},
journal={Food Hydrocoll.},
volume={114},
pages={106546},
year={2021},
publisher={Elsevier}
}

@article{franco2021,
title={Effect of liquid elasticity on the behaviour of high-speed focused jets},
author={Franco-G{\'o}mez, A and Onuki, H and Yokoyama, Y and Nagatsu, Y},
journal={Exp. Fluids},
volume={62},
pages={1--15},
year={2021},
publisher={Springer}
}

@article{bazilevskii2005,
title={Dynamics and breakup of pulse microjets of polymeric liquids},
author={Bazilevskii, AV and Meyer, JD and Rozhkov, AN},
journal={Fluid Dyn.},
volume={40},
number={3},
pages={376--392},
year={2005},
publisher={Springer}
}

@article{hoath2009,
title={Links between ink rheology, drop-on-demand jet formation, and printability},
author={Hoath, SD and Hutchings, IM and Martin, GD and Tuladhar, TR and Mackley, MR and Vadillo, D},
journal={J. Imaging Sci. Technol.},
volume={53},
number={4},
pages={041208},
year={2009}
}

@article{morrison2010,
title={Viscoelasticity in inkjet printing},
author={Morrison, Neil F and Harlen, Oliver G},
journal={Rheol. Acta},
volume={49},
pages={619--632},
year={2010},
publisher={Springer}
}

@article{antonopoulou2020,
title={Jetting behavior in drop-on-demand printing: Laboratory experiments and numerical simulations},
author={Antonopoulou, E and Harlen, OG and Walkley, MA and Kapur, N},
journal={Phys. Rev. Fluids},
volume={5},
number={4},
pages={043603},
year={2020},
publisher={APS}
}

@article{zhao2021,
title={Drop-on-demand ({DOD}) inkjet dynamics of printing viscoelastic conductive ink},
author={Zhao, Dengke and Zhou, Hongzhao and Wang, Yifan and Yin, Jun and Huang, Yong},
journal={Addit. Manuf.},
volume={48},
pages={102451},
year={2021},
publisher={Elsevier}
}

@article{sarban2011,
title={Dynamic electromechanical modeling of dielectric elastomer actuators with metallic electrodes},
author={Sarban, Rahimullah and Lassen, Benny and Willatzen, Morten},
journal={IEEE/ASME Trans. Mechatron.},
volume={17},
number={5},
pages={960--967},
year={2011},
publisher={IEEE}
}

@article{clasen2006,
title={The beads-on-string structure of viscoelastic threads},
author={Clasen, Christian and Eggers, Jens and Fontelos, Marco A and Li, Jie and McKinley, Gareth H},
journal={J. Fluid Mech.},
volume={556},
pages={283--308},
year={2006},
publisher={Cambridge University Press}
}

@article{christanti2002,
title={Effect of fluid relaxation time of dilute polymer solutions on jet breakup due to a forced disturbance},
author={Christanti, Yenny and Walker, Lynn M},
journal={J. Rheol.},
volume={46},
number={3},
pages={733--748},
year={2002},
publisher={The Society of Rheology}
}

@article{torres2022,
title={Flow regimes of a single pulse of inelastic and elastic liquids from a round capillary tube},
author={Torres, Pedro and Fonte, Cl{\'a}udio P},
journal={Chem. Eng. Sci.},
volume={248},
pages={117245},
year={2022},
publisher={Elsevier}
}

@article{turkoz2018,
title={Impulsively induced jets from viscoelastic films for high-resolution printing},
author={Turkoz, Emre and Perazzo, Antonio and Kim, Hyoungsoo and Stone, Howard A and Arnold, Craig B},
journal={Phys. Rev. Lett.},
volume={120},
number={7},
pages={074501},
year={2018},
publisher={APS}
}

@article{yukisada2018,
title={Enhancement of focused liquid jets by surface bubbles},
author={Yukisada, Ryosuke and Kiyama, Akihito and Zhang, Xuehua and Tagawa, Yoshiyuki},
journal={Langmuir},
volume={34},
number={14},
pages={4234--4240},
year={2018},
publisher={ACS Publications}
}

@article{delrot2018,
title={Depth-controlled laser-induced jet injection for direct three-dimensional liquid delivery},
author={Delrot, Paul and Hauser, Sylvain P and Krizek, Jan and Moser, Christophe},
journal={Appl. Phys. A},
volume={124},
pages={1--8},
year={2018},
publisher={Springer}
}

@article{kamamoto2021ouzo,
title={Ouzo column under impact: Formation of emulsion jet and oil-lubricated droplet},
author={Kamamoto, Kyota and Kiyama, Akihito and Tagawa, Yoshiyuki and Zhang, Xuehua},
journal={Langmuir},
volume={37},
number={6},
pages={2056--2064},
year={2021},
publisher={ACS Publications}
}

@article{van2014,
title={Velocity profile inside piezoacoustic inkjet droplets in flight: Comparison between experiment and numerical simulation},
author={Van der Bos, Arjan and van der Meulen, Mark-Jan and Driessen, Theo and van den Berg, Marc and Reinten, Hans and Wijshoff, Herman and Versluis, Michel and Lohse, Detlef},
journal={Phys. Rev. Appl.},
volume={1},
number={1},
pages={014004},
year={2014},
publisher={APS}
}

@article{miyazaki2021,
title={Dynamic mechanical interaction between injection liquid and human tissue simulant induced by needle-free injection of a highly focused microjet},
author={Miyazaki, Yuta and Usawa, Masashi and Kawai, Shuma and Yee, Jingzu and Muto, Masakazu and Tagawa, Yoshiyuki},
journal={Sci. Rep.},
volume={11},
number={1},
pages={14544},
year={2021},
publisher={Nature Publishing Group UK London}
}

@article{chilcott1988,
title={Creeping flow of dilute polymer solutions past cylinders and spheres},
author={Chilcott, MD and Rallison, John M},
journal={J. Non-Newtonian Fluid Mech.},
volume={29},
pages={381--432},
year={1988},
publisher={Elsevier}
}

@article{hoath2012,
title={Jetting behavior of polymer solutions in drop-on-demand inkjet printing},
author={Hoath, Stephen D and Harlen, Oliver G and Hutchings, Ian M},
journal={J. Rheol.},
volume={56},
number={5},
pages={1109--1127},
year={2012},
publisher={AIP Publishing}
}

@article{mcilroy2013,
title={Modelling the jetting of dilute polymer solutions in drop-on-demand inkjet printing},
author={McIlroy, Claire and Harlen, OG and Morrison, NF},
journal={J. Non-Newtonian Fluid Mech.},
volume={201},
pages={17--28},
year={2013},
publisher={Elsevier}
}

@article{cheng2022,
title={Centrifugal multimaterial {3D} printing of multifunctional heterogeneous objects},
author={Cheng, Jianxiang and Wang, Rong and Sun, Zechu and Liu, Qingjiang and He, Xiangnan and Li, Honggeng and Ye, Haitao and Yang, Xingxin and Wei, Xinfeng and Li, Zhenqing and others},
journal={Nat. Commun.},
volume={13},
number={1},
pages={7931},
year={2022},
publisher={Nature Publishing Group UK London}
}

@article{noor2019,
title={{3D} printing of personalized thick and perfusable cardiac patches and hearts},
author={Noor, Nadav and Shapira, Assaf and Edri, Reuven and Gal, Idan and Wertheim, Lior and Dvir, Tal},
journal={Adv. Sci.},
volume={6},
number={11},
pages={1900344},
year={2019},
publisher={Wiley Online Library}
}

@article{demei2022,
title={{3D} food printing: Controlling characteristics and improving technological effect during food processing},
author={Demei, Kong and Zhang, Min and Phuhongsung, Pattarapon and Mujumdar, Arun S},
journal={Food Res. Int.},
volume={156},
pages={111120},
year={2022},
publisher={Elsevier}
}

@article{mantihal2020,
title={{3D} food printing of as the new way of preparing food: A review},
author={Mantihal, Sylvester and Kobun, Rovina and Lee, Boon-Beng},
journal={Int. J. Gastron. Food Sci.},
volume={22},
pages={100260},
year={2020},
publisher={Elsevier}
}

@article{yokoyama2023,
title={Integrated photoelasticity in a soft material: Phase retardation, azimuthal angle, and stress-optic coefficient},
author={Yokoyama, Yuto and Mitchell, Benjamin R and Nassiri, Ali and Kinsey, Brad L and Korkolis, Yannis P and Tagawa, Yoshiyuki},
journal={Opt. Lasers Eng.},
volume={161},
pages={107335},
year={2023},
publisher={Elsevier}
}

@article{bhat2010,
  title={Formation of beads-on-a-string structures during break-up of viscoelastic filaments},
  author={Bhat, Pradeep P and Appathurai, Santosh and Harris, Michael T and Pasquali, Matteo and McKinley, Gareth H and Basaran, Osman A},
  journal={Nat. Phys.},
  volume={6},
  number={8},
  pages={625--631},
  year={2010},
  publisher={Nature Publishing Group UK London}
}

@article{turkoz2018-2,
  title={Axisymmetric simulation of viscoelastic filament thinning with the {Oldroyd-B} model},
  author={Turkoz, Emre and Lopez-Herrera, Jose M and Eggers, Jens and Arnold, Craig B and Deike, Luc},
  journal={J. Fluid Mech.},
  volume={851},
  pages={R2},
  year={2018},
  publisher={Cambridge University Press}
}

@article{turkoz2021,
  title={Simulation of impulsively induced viscoelastic jets using the {Oldroyd-B} model},
  author={Turkoz, Emre and Stone, Howard A and Arnold, Craig B and Deike, Luc},
  journal={J. Fluid Mech.},
  volume={911},
  pages={A14},
  year={2021},
  publisher={Cambridge University Press}
}

@article{kurihara2025,
  title={Pressure fluctuations of liquids under short-time acceleration},
  author={Kurihara, Chihiro and Kiyama, Akihito and Tagawa, Yoshiyuki},
  journal={J. Fluid Mech.},
  volume={1003},
  pages={A20},
  year={2025},
  publisher={Cambridge University Press}
}

@article{ng2024,
  title={Jetting-based bioprinting: Process, dispense physics, and applications},
  author={Ng, Wei Long and Shkolnikov, Viktor},
  journal={Bio-Des. Manuf.},
  volume={7},
  number={5},
  pages={771--799},
  year={2024},
  publisher={Springer}
}

@article{subbotin2023,
  title={The elasticity of polymer melts and solutions in shear and extension flows},
  author={Subbotin, Andrey V and Malkin, Alexander Ya and Kulichikhin, Valery G},
  journal={Polymers},
  volume={15},
  number={4},
  pages={1051},
  year={2023},
  publisher={MDPI}
}

@article{nakamine2024,
  title={Flow birefringence of cellulose nanocrystal suspensions in three-dimensional flow fields: Revisiting the stress-optic law},
  author={Nakamine, Kento and Yokoyama, Yuto and Worby, William Kai Alexander and Muto, Masakazu and Tagawa, Yoshiyuki},
  journal={Cellulose},
  volume={31},
  number={12},
  pages={7405--7420},
  year={2024},
  publisher={Springer}
}

@article{worby2024,
  title={Examination of flow birefringence induced by the shear components along the optical axis using a parallel-plate-type rheometer},
  author={Worby, William Kai Alexander and Nakamine, Kento and Yokoyama, Yuto and Muto, Masakazu and Tagawa, Yoshiyuki},
  journal={Sci. Rep.},
  volume={14},
  number={1},
  pages={21931},
  year={2024},
  publisher={Nature Publishing Group UK London}
}

@article{dixit2025,
  title={Viscoelastic {Worthington} jets and droplets produced by bursting bubbles},
  author={Dixit, Ayush and Oratis, Alexandros and Zinelis, Konstantinos and Lohse, Detlef and Sanjay, Vatsal},
  journal={J. Fluid Mech.},
  volume={1010},
  pages={A2},
  year={2025},
  publisher={Cambridge University Press}
}

@article{gaillard2025,
  title={When does the elastic regime begin in viscoelastic pinch-off?},
  author={Gaillard, Antoine and Herrada, MA and Deblais, Antoine and Van Poelgeest, C and Laruelle, L and Eggers, J and Bonn, Daniel},
  journal={J. Fluid Mech.},
  volume={1005},
  pages={A10},
  year={2025},
  publisher={Cambridge University Press}
}

@article{oratis2025,
  title={Viscoelastic lubrication of a submerged cylinder sliding down an incline},
  author={Oratis, Alexandros T and van den Berg, Kai and Bertin, Vincent and Snoeijer, Jacco H},
  journal={Europhys. Lett.},
  volume={149},
  number={6},
  pages={63002},
  year={2025},
  publisher={IOP Publishing}
}

@article{balasubramanian2024,
  title={Bursting bubble in an elastoviscoplastic medium},
  author={Balasubramanian, Arivazhagan G and Sanjay, Vatsal and Jalaal, Maziyar and Vinuesa, Ricardo and Tammisola, Outi},
  journal={J. Fluid Mech.},
  volume={1001},
  pages={A9},
  year={2024},
  publisher={Cambridge University Press}
}

@article{sari2024,
  title={The effect of fluid viscoelasticity in soft lubrication},
  author={Sari, MH and Putignano, Carmine and Carbone, Giuseppe and Biancofiore, Luca},
  journal={Tribol. Int.},
  volume={195},
  pages={109578},
  year={2024},
  publisher={Elsevier}
}

@article{wang2024,
  title={Experimental and numerical study on dynamics of viscoelastic liquid cone in flow focusing},
  author={Wang, Ming and Mu, Kai and Zhao, Chengxi and Wu, Yanfeng and Xu, Wenshuai and He, Xiuli and Si, Ting},
  journal={Phys. Fluids},
  volume={36},
  number={9},
  year={2024},
  publisher={AIP Publishing}
}

\end{document}